\documentclass[epj]{svjour}

\usepackage{graphics}
\begin{document}
\title{High-order harmonic generation at high laser intensities beyond the tunnel regime}

\author{J. A. P{\'e}rez-Hern{\'a}ndez\inst{1}
\and M. F. Ciappina\inst{2}\thanks{\emph{Present address:} Max-Planck-Institut f\"ur Quantenoptik, Hans-Kopfermann-Strasse 1, D-85748 Garching, Germany}  \and M. Lewenstein\inst{3,4} \and A. Za\"ir\inst{5} \and L. Roso\inst{1}}
     
%
\institute{Centro de L\'aseres Pulsados (CLPU), Parque Cient\'{\i}fico, 37185 Villamayor, Salamanca, Spain \and Department of Physics, Auburn University, Auburn, Alabama 36849, USA \and  ICFO-Institut de Ci\'encies Fot\'oniques, Mediterranean Technology Park, 08860 Castelldefels (Barcelona), Spain \and ICREA-Instituci\'o Catalana de Recerca i Estudis Avan\c{c}ats, Lluis Companys 23, 08010 Barcelona, Spain \and Department of Physics, Blackett Laboratory, Imperial College London, London SW7 2AZ, United Kingdom
}
\date{Received: date / Revised version: date}
%
\abstract{We present studies of high-order harmonic generation (HHG) at laser intensities well above saturation. We use driving laser pulses which present a particular electron dynamics in the turn-on stage. Our results predict an increasing on the harmonic yield, after an initial dropping, when the laser intensity is increased. This fact contradicts the general belief of a progressive degradation of the harmonic emission at ultrahigh intensities. We have identified a particular set of trajectories which emerges in the turn-on stage of these singular laser pulses, responsible of the unexpected growth on the harmonic efficiency at this high intensity regime. Our study combines two complementary approaches: classical analysis and full quantum mechanical calculations resulting from the numerical integration of the 3-dimensional time-dependent Schr{\"o}dinger equation complemented with the time-frequency analysis.
} 

\maketitle
\section{Introduction}
\label{intro}
High-order harmonics generation (HHG) provides a direct and elegant route to achieve coherent radiation in the range of soft x-rays using table-top laser sources~\cite{seres,tenio}. In the HHG spectrum the harmonic cut-off determines the highest energies possible to achieve for each specific input laser pulse interacting with a target atom, molecule or ion~\cite{corkum,lewenstein,kulander}. In this way a large number of photons is desirable at the output but, at the same time, a high harmonic efficiency is also required. In other case, this process it turns to be inefficient. 
Two possible ways to achieve long harmonic cut-offs are in order. One relies in the increment of the input laser intensity and the other one takes advantage of the HHG scaling with the input laser wavelength. On the one hand, it is well known that when the laser wavelength $\lambda$ is increased, the cut-off increases quadratically with it, but at the same time the single atom efficiency decreases exponentially as $\lambda^{-5.5}$~\cite{tate,schiessl}. Recent works have demonstrated that this loss of efficiency can be compensated by optimizing the phase matching conditions using hollow fibers~\cite{tenio}. However, in spite that this method is a promising way, requires broad technology devices, such as Optical Parametric Chirped-pulse Amplification (OPCPA) laser sources and complex experimental set ups in the laboratory. Another theoretical approaches treat to increase the harmonic cut-off by using chirped laser pulses or by mixing different colors~\cite{perfect,perfectexp,carrera,hong,zeng,zhai,oishi,xue}, delaying in time two identical pulse replicas and its ulterior overlapping with the adequate delay~\cite{McD} or inducing a weak spatial inhomogeneity in the interacting field which occurs when a laser pulse is focused in the surrounding of the nanotips, nanoparticles and/or nanoplasmonic antenas~\cite{kim,yavuz,ciappi2012,ciappi2012_bis,tahir2012,ciappi_opt,ciappiAdP,miloAdP,joseprl2013,yavuz2013,ciappicpc2014}. Alternatively the target active medium may be chosen to extend the cut-off, for example using neighboring centers can achieve up to $8U_p+I_p$ frequencies~\cite{Numico98,moreno97,Bandrauk07}, but these methods rely on the precise control of the internuclear separation of a molecule.

The other alternative way of achieving high harmonics cut-offs is by increasing the input laser intensity which produces a cut-off linear scaling with this quantity. But unfortunately this growth is not unlimited and exists a physical limit: the so-called {\em saturation threshold} which depends on the atomic structure of the target atom, molecule or ion. In this way, when the laser intensity is increased beyond this critical value, the harmonic efficiency drastically drops making this process inefficient~\cite{moreno,strelkov}. In other words, from the experimental view point, the saturation causes that the harmonic signal drops some orders of magnitude making it too weak to be experimentally detected. In this work we will expose a tentative approach different from the conventional techniques mentioned above, to achieve high energetic harmonic cut-offs (i.e. high-energy photons). This technique emerges from a detailed study of the intra-cycle electron dynamics which take place in the turn-on stage of a particular kind of driving laser pulses.

The present article is organized as follows. In the next section, Section 2, we describe the two different theoretical approaches used to study HHG generated by Non-Adiabatic Turn-on (NAT) pulses. On the one hand the so-called classical analysis is presented (Section 2.1). On the other hand, in Section 2.2, the full quantum mechanical description is carried out, based on the numerical resolution of the 3-Dimensional Time Dependent Schr{\"o}dinger Equation in the single active electron approximation. Next, in Section 3, we present numerical results mainly focused in the argon atom. Additionally, in order to compare both (classical and quantum) predictions we have also included the time-frequency analysis which complements, in detail, the harmonic emission in the quantum mechanical context, allowing to compare them with the classical results. Finally, in Section 4, we close our contribution with our conclusions and a brief outlook. Atomic units are used throughout the article unless otherwise stated.

\section{Theory}
\label{sec:1}

\subsection{Classical analysis}
\label{subsec:1}

A qualitative description of high-order harmonic generation can be retrieved by using the three step model \cite{corkum,lewenstein,kulander}. The classical trajectory associated to the corresponding electron ionized by the laser field can be computed by integrating the Newton-Lorentz equation in the dipole approximation. In Fig.~1 we plot the rescattering kinetic energy as a function of the
ionization time (grey circles) and as a function of the recombination time (black triangles). Figure 1(a)
represents a constant envelope laser field, meanwhile Fig.~1(b) and Fig.~1(c) are for a few cycle laser pulse (see below).  For the case of a constant laser field, Fig.~1(a), we can observe the most energetic trajectories correspond to
electrons ionized near the peak pulse, that recollides with the parent ion at approximately
3/4 of a cycle later. These kind of trajectories correspond to the so called {\em long} and {\em short} \cite{corkum,lewenstein,kulander} which explain the cut-off law associated to the harmonic plateau. Note also that the non returning electrons can achieve values around $I_p+8U_p$ but in this case do not contribute to generate HHG \cite{jose_ks}

On the other hand, Fig.~1(b) and Fig.~1(c) correspond to a few cycles pulses (1.5 cycles FWHM) with
the same envelope and different CEP, with the explicit form 
\begin{equation}
\label{pulse}
E(t)=E_0\sin^{2}\left(\frac{\omega t}{2N}\right)\sin(\omega t+\phi)
\end{equation}

%

if $ 0 <  t  < {NT} $ and $E(t)=0$ in otherwise. $T=2\pi/\omega$, $E_0$ is the field peak amplitude,  $\phi$ is the so-called carrier-envelope phase (CEP) and N is the total number of cycles. In this case N=3 and $\lambda=800nm$, consequently one laser cycle corresponds to $2.6fs$.
Note that in the case of $\phi=0$ (Fig.~1(b)) a new kind of trajectories emerges in
the turn-on, labeled as a NAT (Non Adiabatic Turn-on). This kind of trajectories are different from the usual long and short trajectories. In
the following we will consider $\phi=0$, unless otherwise is specified. Through the analysis of Fig.~1(b) it is possible to identify the long and short type of trajectories,
i.e. with the most energetic recollisions corresponding to electrons ionized near to the
peak field. The second class of trajectories, NAT, correspond to the electrons ionized
during the early turn-on. 

\begin{figure}[h]
\begin{center}
\resizebox{0.40\textwidth}{!}{\includegraphics{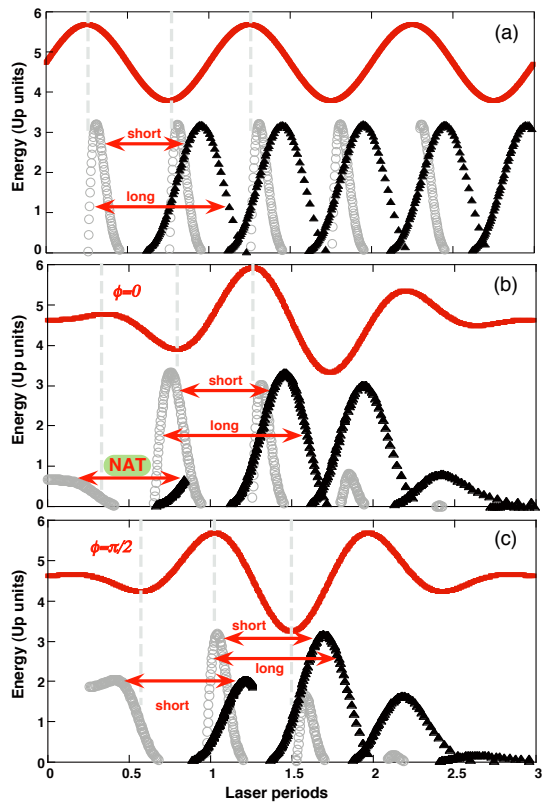}}
\caption{Recombination energies of ionized electrons as a function
of the ionization time (gray circles) and recombination time (black triangles), for three
different laser pulses (in red): (a) constant envelope, and (b) and (c) as described in Eq.~(\ref{pulse})
(1) with (b) $\phi=0$ and (c) $\phi=\pi/2$. Note that for the case $\phi=0$ (panel b), a new set of trajectories emerges in the turn-on, labeled as NAT.}
\label{fig:1}       
\end{center}
\end{figure}

A more detailed view of the Fig.~1(b) is plotted
in Fig.~2. Here, it is shown that these kind of trajectories follow the opposite trend:
the most energetic rescatterings originate from electrons ionized almost a quarter cycle
before the first field maximum. The emergence of NAT trajectories is associated with
the fast turn-on of the laser pulse. Note that in this case the recombination energies are close to $0.5U_p$, $U_p$ being $U_p=E_0^{2}/4\omega^{2}$ the ponderomotive energy. In spite that these energy values are low compared to the conventional $3.17U_p$, as we will see, it turns important when the laser intensity achieves values far beyond the saturation threshold where for usual pulses with adiabatic turn-on, the harmonic yield is degraded~\cite{moreno,strelkov}. 

As we will see, this behavior is related to the decay in efficiency of the harmonic radiation generated by the
conventional short and long trajectories. This decay is a direct consequence of the fast depopulation of the ground state during the excursion of the electron in the continuum. On this manner, at the time of rescattering, the ground state is totally depopulated and the dipole amplitude is relatively small resulting in a lack of harmonic efficiency produced by long and short trajectories.
\begin{figure}[h]
\begin{center}
\resizebox{0.40\textwidth}{!}{\includegraphics{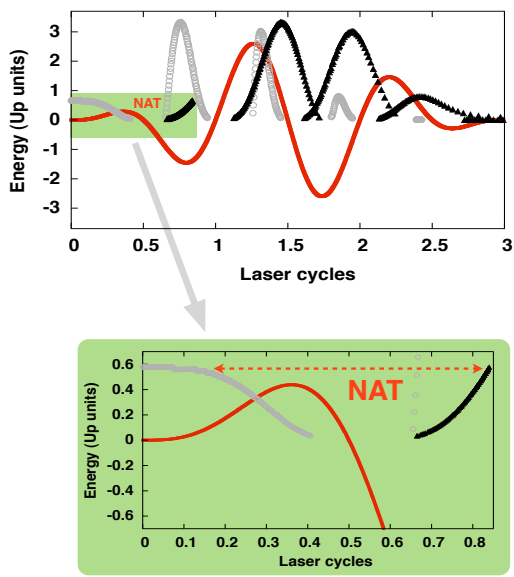}}
\caption{Detailed view of Fig.~1(b) $(\phi=0)$ in which is highlighted the NAT
trajectories emerging in the turn-on. Note that in this case electrons ionized in the early turn-on recombine during the first cycle, before the atom was totally depleted. This energies are close to $0.5U_p$, $U_p$ being $U_p=E_0^{2}/4\omega^{2}$ the ponderomotive energy. In spite that these energy values are low, they turns important when the laser intensity achieves values far beyond the saturation}
\label{fig:2}       
\end{center}
\end{figure}
It is worth to mention that the NAT trajectories also emerges in pulses with other envelopes. In particular we have reproduced comparable results by using pulses with Gaussian envelopes and similar turn-on than the pulses described by Eq. (1). 
\subsection{Quantum mechanical approach}
\label{subsec:2}
In order to make a quantitative evaluation of the relative contributions to the harmonic
spectrum of the different trajectories involved (long, short and NAT) we will calculate the absolute value of the complex dipole amplitude (assuming constant the transition matrix elements)
\begin{equation}
\label{dipole}
|d(t)|\propto|a_0^{*}(t)||a_{\mathbf{v}}(t)|
\end{equation}
where $a_0(t)$ is the transition amplitude of the ground state and $a_{\mathbf{v}}(t)$ is the transition
amplitude of the free electron state with velocity $\mathbf{v}$ at the time of rescattering
$t$. The values for the transition amplitudes are extracted from the results of the exact
numerical integration of the 3-Dimensional time-dependent Schr\"odinger equation (3D-TDSE): $|a_0(t)|$
is found by projecting the total wavefunction on the ground state, and $|a_{\mathbf{v}}(t)|$ is estimated
computing the ground-state depletion during a small time-window around the
corresponding ionization time $t_0$ (i.e. the initial time of the trajectory associated to
the rescattering at time $t$). Specifically,
\begin{equation}
|a_v(t)|^2 \simeq \left . {d \over dt} |a_0|^2 \right |_{t_0} \Delta t
\end{equation}
with $\Delta t$ being a small time interval, whose particular value is not important for the
relative comparison between different trajectories, as long as it is kept unchanged. The
values of the ionization and rescattering times ($t_0$ and $t$) for a particular trajectory are
extracted from the classical analysis of Fig.~1. This allow us to associate each
pair $(t_0, t)$ to a well-defined trajectory of the NAT or long and short. In order to compare the harmonic efficiency at different regimes of laser intensity, we compute the yield at a fixed energy: $73$ eV, which corresponds to the harmonic cut-off in the case of the threshold of saturation (see Fig.~4).
\begin{figure}[h]
\begin{center}
\resizebox{0.40\textwidth}{!}{\includegraphics{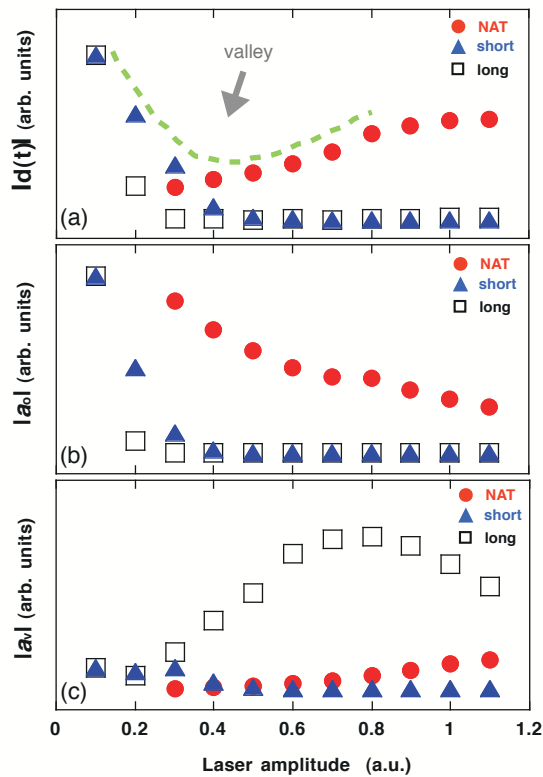}}
\caption{Estimations of (a) the relative contributions to the harmonic yield computed at
$73$ eV, which corresponds to the cut-off energy in the threshold of saturation case (see Fig.~\ref{fig:4}). (b) the probability amplitude of the ground state, and (c) the corresponding
continuum state at the moment of recollision for the sets of trajectories highlighted
in Fig.~1(b). Contributions of NAT trajectories are shown as red circles, long trajectories
as open squares and short as blue triangles. The dashed line plotted in (a) represents a sketch
of the resulting valley formed by the total yield (sum of the above contributions).
The laser amplitudes E are given in atomic units (a.u.), corresponding to intensities
$E^2\times3.51\times10^{16}$ W/cm$^2$.}
\label{fig:3}       
\end{center}
\end{figure}
On this way, for each laser intensity, we use Fig.~1 to determine the ionization time
$t_0$, and the rescattering time $t$ corresponding to the electronic trajectories with recollision
kinetic energy $73$ eV$-I_p$. The results for the estimations of the harmonic efficiencies at $73$ eV using
different laser intensities is shown in Fig.~3(a), for the few cycle pulse considered
in Fig.~2(b). Long (open squares) and short (blue triangles) curves in Fig.~3(a) show a decaying behavior with
the increment of the field amplitude, which is related to the degradation of the harmonic
generation by these type of trajectories. The reason behind this behavior can be found in the
analysis of the probability amplitudes which conform the dipole transition (see Eq.~(\ref{dipole})). Fig.~3(a) and Fig.~3(c) show the probability amplitudes of the ground and continuum states at rescattering, $|a_0(t)|$ and $|a_{\mathbf{v}}(t)|$. As it is apparent, the decrease
in the harmonic efficiency radiated by long and short trajectories is related to the fast ionization of the ground state for intensities above saturation and, therefore, with the decrease of $|a_0(t)|$. Despite of the fact that depletion of the ground
state population increases the population of electrons in the continuum, i.e. $|a_{\mathbf{v}}(t)|$ increases
when $|a_0(t)|$ decreases, the product of both amplitudes has a net decrease and the
efficiency of the dipole transition falls down. In contrast, for the case of NAT trajectories,
the behavior is the opposite: as they are originated at the first part of the turn-on,
the ionization is relatively moderate, even in the case of field amplitudes one order of magnitude
above saturation (i.e. intensities two orders of magnitude above saturation). At
rescattering, therefore, there is still a significant population in the ground state, and
the product of probability amplitudes does not vanish. Therefore, the dipole strength
is found to increase gradually with the laser amplitude. Consequently, when the laser amplitude is increased (Fig.~3(a)) the harmonic yield tends to form a valley: firstly, a decrease connected with the degradation of the efficiency of the short and long
trajectories, followed by an increase as the efficiency of the NAT trajectories becomes
the relevant contribution to the dipole spectrum. NAT trajectories will eventually are
degraded for ultraintense fields well above the atomic unit ($3.51\times10^{16}$ W/cm$^2$), however
for these intensities we should expect also a decay connected with the breaking
of the dipole approximation and the associated drift of the electron trajectories away
from the ion due to the interaction with the magnetic field~\cite{javi}.

\section{Results}
\label{sec:3}
In order to confirm the asseverations presented in the previous section, Fig.~4 shows the harmonic spectrum at laser intensities corresponding
to threshold of saturation (solid gray), saturation (solid blue) and deep saturation (solid red, green and black) regimes, extracted from the
exact integration of the 3D-TDSE in Argon ($I_p=-15.7$ eV). For this goal we have implemented in our 3D TDSE code the atomic potential extracted from~\cite{tong} which describes accurately the Argon atom within the context of the single active electron (SAE) approximation . A similar study has been carried out in Hydrogen as is reported in~\cite{josenat}. Here we have focused in Argon due to this noble gas is one of the species more commonly used in the HHG experiments.

Below and at the saturation threshold ($I \le 3.51\times10^{14}$ W/cm$^2$) the increase in laser intensity does not affect strongly the
harmonic yield, although it extends the harmonic plateau accordingly to the semiclassical cut-off
law $I_p+3.17U_p$. Above saturation the harmonic yield begins to decrease until a
minimum is reached at $I\approx5.6\times10^{15}$ W/cm$^2$, corresponding to the bottom of the valley
structure sketched in Fig.~3(a). For even higher intensities the harmonic yield increases as
a consequence of the emergence of the contribution of the NAT-type trajectories to the
radiation spectrum.

Regarding to the behavior with the CEP, when a few cycle input laser pulse as described by Eq.~(\ref{pulse}) is used, the interaction of NAT pulses with matter is strongly dependent on the CEP, $\phi$. Therefore, if the CEP of the ultra-short pulse changes from
$\phi=0$ to $\pi/2$ the Fig.~1(c) shows the energy diagram for the trajectories. In this
case, the only relevant trajectories for harmonic generation are the conventional short
and long type, consequently NAT trajectories are not useful, as the electron excursion
is too short to acquire energy relevant for HHG before the total depopulation of the
atom. This distinct behavior is also confirmed by the 3D-TDSE simulations. However, it is important to
stress out that the NAT-pulses support small CEP fluctuations $(\pm0.1\pi\mathrm{rad})$ typical from
the CEP-stabilized laser sources. This behavior is shown in Fig.~4 by green and black solid lines.

Additionally, it has been demonstrated in~\cite{libro} that for longer pulses and identical turn-on stage that those of Fig.~1(b), the NAT effect is preserved providing similar gains in the harmonic yield, and consequently, the role of the CEP is less relevant.  
\begin{figure}[h]
\begin{center}
\resizebox{0.45\textwidth}{!}{\includegraphics{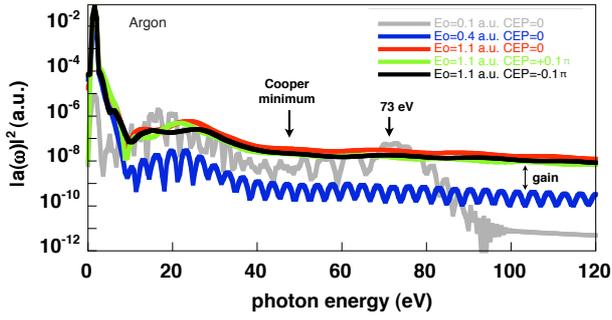}}
\caption{Spectra resulting from the exact integration of the 3-dimensional time
dependent Schr\"odinger equation (3D-TDSE) in Argon at $\lambda=800$ nm for the laser pulse described
in Eq.~(\ref{pulse}) with $\phi=0$ (Fig.~1(b)), for different values of the laser intensity: threshold
of saturation ($I = 3.51\times10^{14}$ W/cm$^2$, solid gray line), saturation ($I = 5.6\times10^{15}$ W/cm$^2$, solid blue line) and deep saturation ($I = 4.2\times10^{16}$ W/cm$^2$, solid red, green and black lines). In particular, green and black lines represents small fluctuations $(\pm0.1\pi\mathrm{rad})$ in the CEP in the case of the deep saturation intensity}
\label{fig:4}       
\end{center}
\end{figure}
Fig.~5 represents the full 3D TDSE spectrum in Argon for the deep saturation case $I = 4.2\times10^{16}$ W/cm$^2$ at $800$ nm and $\phi=0$. Note that here the cut-off achieved ($I_p+0.5U_p$) confirms the classical prediction of Fig.~1(b) and exceed the water window region reaching values of photon energies around the keV.
\begin{figure}[h]
\begin{center}
\resizebox{0.50\textwidth}{!}{\includegraphics{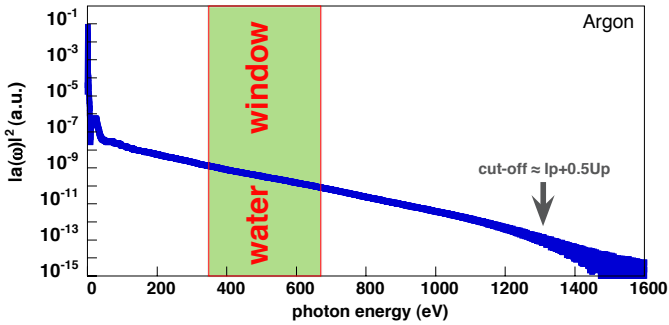}}
\caption{Full 3D TDSE spectrum in Argon at $\lambda=800$ nm for the case of deep saturation regime ($I = 4.2\times10^{16}$ W/cm$^2$). Note that the water window region is exceeded.}
\label{fig:5}       
\end{center}
\end{figure}

In order to study the NAT harmonic emission within the full quantum context and compare with the classical predictions carried out in Section 2.1, we have
performed the time-frequency analysis of the dipole acceleration extracted from the
exact integration of the 3D-TDSE. The time-frequency analysis provides the extraction
of the information about the times and efficiencies at which an specific wavelength is
radiated. In the Fig.~6 is plotted the time-frequency analysis for the three
intensity values presented in the Fig.~4, corresponding to the NAT-type pulse plotted in
Fig.~1(b). In order to compare with the classical predictions we also have superimposed
over the time-frequency results, the rescattering energies computed from the classical
trajectories labeled in Fig.~1(b). By inspection of the three panels of Fig.~6 it is easy
to see that in the case of threshold of saturation Fig.~6(a) the conventional short and long trajectories
are responsible of the harmonic generation according to the cut-off law $3.17U_p$. When the intensity exceeds the threshold of saturation, the harmonic efficiency produced by long and short trajectories is gradually
degraded (Fig.~6(b)). Note that, at this intensity regime, there are evidences that
the NAT trajectories turn to be important. Fig.~6(b) shows the deep saturation case where
only NAT trajectories survive due to the fact that the atom is being completely depleted, except in
approximately the first half period of the driving laser pulse. For this reason, short and
long trajectories are not involved in the harmonic generation at these high intensity
values and obviously there are no interferences between both. It is important to point out that the time-frequency analysis of Fig.~6(c) confirms that electrons recombine with energies around $0.5U_p$ in agreement with the classical prediction reported in Fig.~1(b). 

According to these results, in deep saturation regime, only NAT trajectories are
responsible of the harmonic radiation. This is the cause of the absence of modulations
in the harmonic spectrum of the deep saturation case plotted in Fig.~5. This fact makes to suspect that 
 an narrow and intense attosecond pulse could be obtained from the totality of the plateau in the deep
saturation spectrum.

\begin{figure}[h]
\begin{center}
\resizebox{0.45\textwidth}{!}{\includegraphics{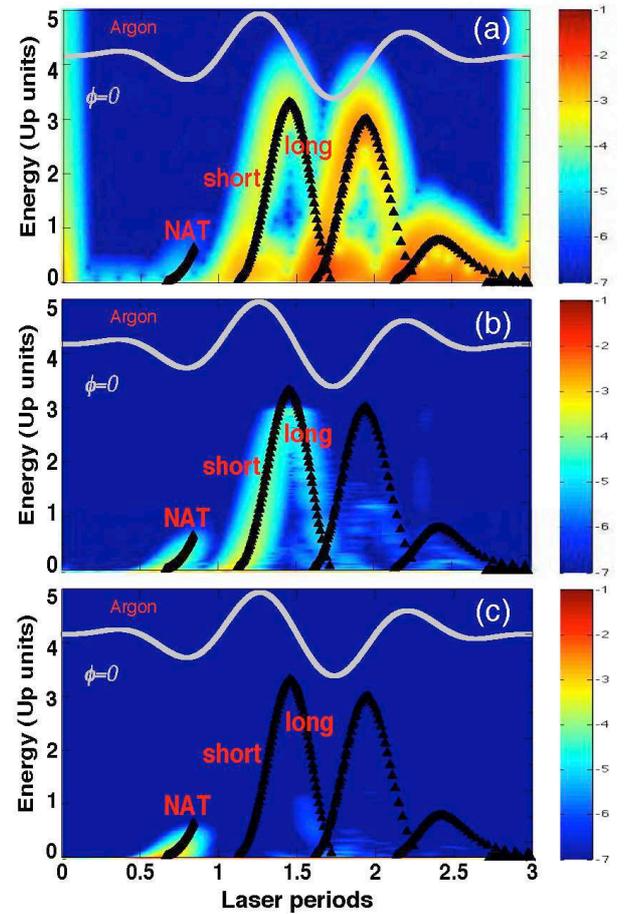}}
\caption{Time-frequency analysis of dipole acceleration extracted from the full computation of the 3D TDSE in Argon, and (superimposed) classical rescattering energies of electrons as a function of the recombination time (black triangles) for the three laser intensities shown in Fig.~\ref{fig:4}. (a) threshold of saturation, (b) saturation and (c) deep saturation. The driving laser pulse is plotted in solid gray lines.}
\label{fig:5}       
\end{center}
\end{figure}

Ê 
To conclude it is important to mention the role of the propagation effects. The 
free electron plasma produced by the high ionization rate induces a strong 
dephasing. This effect can degrade the macroscopic harmonic yield. However, 
previous experimental works, see for instance ~\cite{seres04} and references 
there in, where high laser intensities are used, suggest that the effect known as 
nonadiabatic self-phase matching (NSPM) can be used to compensate the 
dephasing produced by the electron plasma, improving the phase matching. This 
fact makes that the harmonic signal emitted from individual atoms can grow 
coherently in the macroscopic interaction medium.

\section{Conclusions and Outlook}
\label{sec:4}

Summarizing, we have theoretically demonstrated that NAT-type
input laser pulses are robust to the saturation effects and, therefore,
play the fundamental role in harmonic generation at intensities far beyond the tunnel regimes. This fact
contradicts the general belief of a progressive degradation of the harmonic efficiency
due to the effect of the barrier suppression, based on the experience with conventional pulses with
smoother turn-on. 

The potentiality of NAT-type laser pulses relies on to achieve high harmonics cut-offs, reaching values of coherent
radiation in the range of keV from interacting NAT-laser pulses centered in $800$ nm.
Also is worth to mention the possibility to synthesize intense and narrow intense attosecond
pulses from NAT-laser pulses with arbitrary temporal duration.

Finally, we would like to point out that the study presented in this
contribution together with the results previously published in~\cite{josenat,libro}, both constitute as far as we know, the
first theoretical work in the literature focused on to identify the physical mechanism
responsible of this unusual growing of the harmonic yield that occurs (under particular
conditions) when the laser intensity is increased beyond the saturation threshold of atoms and molecules.

\begin{acknowledgement}
We acknowledge the financial support of the MICINN projects (FIS2008-00784 TOQATA, FIS2008-06368-C02-01, and FIS2010-12834), ERC Advanced Grant QUAGATUA and OSYRIS, the Alexander von Humboldt Foundation (M.L.), and the DFG Cluster of Excellence Munich Center for Advanced Photonics. This research has been partially supported by Fundaci\`o Privada Cellex. J.A.P.-H. and L. Roso acknowledge support from Laserlab-Europe (Grant No. EU-FP7 284464). A. Za\"ir acknowledges support from EPSRC Grant No. EP/J002348/1 and Royal Society International Exchange Scheme 2012 Grant No. IE120539. This work was made possible in part by a grant of high performance computing resources and technical support from the Alabama Supercomputer Authority.
\end{acknowledgement}

%

%
%
%

\end{document}